\documentclass[floatfix,epsf,
prb,twocolumn,amsmath,amssymb,showpacs,superscriptaddress]{revtex4}
\usepackage[english]{babel}
\usepackage{graphicx}
\usepackage{subfigure}
\newcommand{\kvec}[2]{\begin{pmatrix} #1 \\ #2 \end{pmatrix} }

\newcommand{\matt}[4]{\begin{pmatrix} #1 & #2 \\ #3 & #4 \end{pmatrix} }
\newcommand{\ve}{\varepsilon}
\begin{document}\setlength{\unitlength}{1mm}
%
\title{
Dirac electrons in a Kronig-Penney potential: \\
dispersion relation and transmission periodic in the strength of the barriers}
%
\author{M. Barbier}
\affiliation{Department of Physics, University of Antwerp, 
Groenenborgerlaan 171, B-2020 Antwerpen, Belgium\\}
\author{P. Vasilopoulos}
\affiliation{Department of Physics, Concordia University, 
7141 Sherbrooke Ouest, Montr\'eal, Quebec, Canada H4B 1R6\\}
\author{F.M. Peeters}
\affiliation{Department of Physics, University of Antwerp, 
Groenenborgerlaan 171, B-2020 Antwerpen, Belgium\\}
\begin{abstract}
    The transmission $T$ and conductance $G$ through one or multiple 
    one-dimensional, $\delta$-function barriers of two-dimensional fermions 
    with a linear energy spectrum are studied. $T$ and  $G$ are {\it periodic} 
    functions of the strength $P$ of the $\delta$-function barrier 
    $V(x,y)/\hbar v_F = P \delta(x)$. 
    The dispersion relation of a Kronig-Penney (KP) model 
    of a superlattice is also a {\it periodic} function of $P$ and causes 
    collimation of an incident electron beam for $P = 2 \pi n$ and $n$ integer. 
    For a KP superlattice with alternating sign of the height of the barriers 
    the Dirac point becomes a Dirac line for 
    $P = (n + 1/2)\pi$.
\end{abstract}
 \maketitle
The study of particle motion in periodic potentials is at the heart of condensed 
matter physics and it  is usually assumed that the energy spectrum is {\it 
parabolic}. One of the earliest examples is the well-known, one-dimensional (1D) 
Kronig-Penney (KP) model, 
\cite{nonrelkp} 
that consists of an infinite succession of  very thin 
($W \to 0$) and very high ($V_0\to \infty$)  barriers, referred to as 
$\delta$-function barriers, but such that their product  $P \propto W V_0$ 
remains constant.  This results in minibands in the electron spectrum.

One may wonder though how such results are modified if the energy is {\it  
linear  in wave vector}. Such a spectrum occurs for relativistic electrons with 
energy $E=\hbar c p>>E_0=m_0c^2$, where  $c$ is the speed of light and $m_0$ the 
bare electron rest mass. Even without neglecting $E_0$ a strict 1D Dirac KP 
model was considered for relativistic quarks \cite{mak}. It is also known that 
electrons can transmit perfectly, upon normal incidence, through arbitrarily 
wide and high barriers, referred to as  Klein paradox or Klein tunneling
\cite{kle}. With the discovery of graphene \cite{nov}, a one-atom thick layer of 
carbon atoms, another system became available in which particles (electrons) 
moving in two dimensions, have a linear spectrum, $E=\hbar v_F k$, 
with ${\vec k}=(k_x, k_y)$ the wave vector. Importantly, carriers in graphene 
behave as chiral, massless  fermions  described by  Dirac's equation without the 
mass term, and move with  the Fermi velocity $v_F\approx c/300$. There is a 
wealth of exceptional properties of graphene, see e.g., Refs \cite{net}.

Because the  carriers in graphene move in two dimensions, tunneling through 
barriers is inherently  two-dimensional (2D) and depends on the direction of the 
incident electron beam even in the absence of a magnetic field. Many authors, 
including ourselves, have studied this tunneling, through single, multiple 
barriers, and  superlattices \cite{kat,per}. Surprisingly, tunneling through 
$\delta$-function barriers has received very little attention\cite{sha} 
and we are not aware of any Dirac KP model for a superlattice in graphene. 
An interesting development was the application of periodic potentials 
to graphene that  turned it into a self-collimating material despite the rather 
unusually high potentials used\cite{parksgs}.

Motivated by all these results and the absence of a systematic treatment of KP 
barriers or superlattices, we study in this work the transmission through such 
structures as well as the dispersion relation of a KP superlattice. 
Although the model may appear a bit unrealistic, since a relatively smooth 
potential is needed to describe the carrier dynamics by the Dirac equation
\cite{shon98}, 
its simplicity is attractive and elucidates certain symmetry properties of 
the spectrum. 
Furthermore one can realize the model by using a potential which is smooth on 
the scale of the atomic distance while remaining imidiate compared to the 
typical electron wavelength. 
The 
unexpected results mentioned in the abstract are in sharp contrast with those 
for carriers with a {\it parabolic} energy spectrum described by the 
Schr\"odinger equation. We will use graphene as an example but the results apply 
to any 2D system with a linear-in-wave-vector spectrum and a two-component 
spinor.

\section{Transmission through a $\delta$-function barrier}
We describe the electronic structure of an infinitely large flat graphene flake 
in single valley approximation 
by the 
zero-mass Dirac equation and consider solutions with energy and wave 
vector near the K point. The Hamiltonian is 
$\mathcal{H} = v_F \vec{\sigma}\cdot \vec{p} + \mathbb{1} V$ with  $\vec{p}$  
the momentum operator and $ \mathbb{1}$ the $2\times 2$ unit matrix. In the 
presence of a  1D potential  $V(x)$ the equation $(\mathcal{H} - E)\psi = 0$ 
admits  solutions of the form $\psi(x) e^{i k_y y}$ where
\begin{equation}\label{eq1_2}
    \psi(x) = \kvec{1}{s e^{i \phi}}e^{i \lambda x}, \quad  \psi(x) = 
    \kvec{1}{-s e^{-i \phi}}e^{-i \lambda x}\,,
\end{equation}
with $\tan \phi = k_y/\lambda$, $s = sign(\ve-u(x))$, 
$\lambda = [(\ve-u(x))^2 - k_y^2]^{1/2}$, $\ve = E/v_F \hbar$, and 
$u(x) = V(x)/v_F \hbar$; $\ve $ and $u(x)$ are in units of inverse length.
As usual, we approximate a  $\delta$-function  barrier with a very thin and very 
high barrier, of width $W ( \rightarrow 0$) and height 
$V_0( \rightarrow \infty$), but keep constant the dimensionless product 
$P = W V_0/ \hbar v_F$ which we call its strength.
\begin{figure}[ht]
  \begin{center}
	\subfigure{\includegraphics[height=2.7cm,width=3cm]{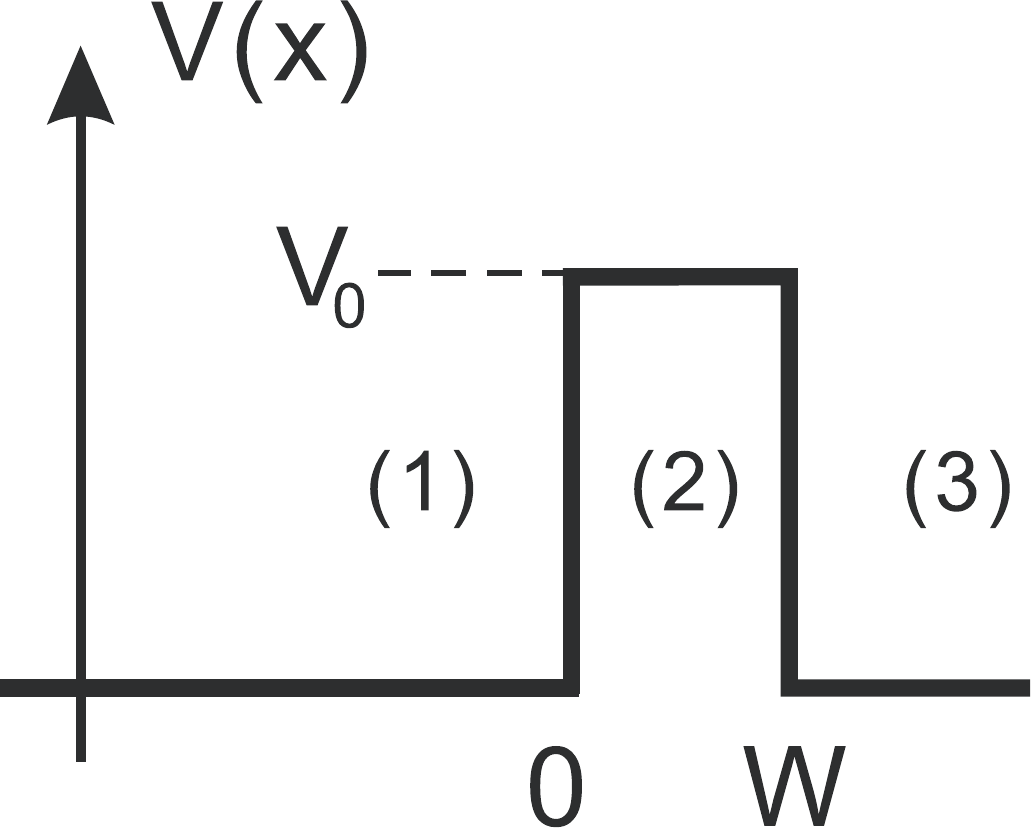}}
    \subfigure{\includegraphics[height=2.7cm,width=3.5cm]{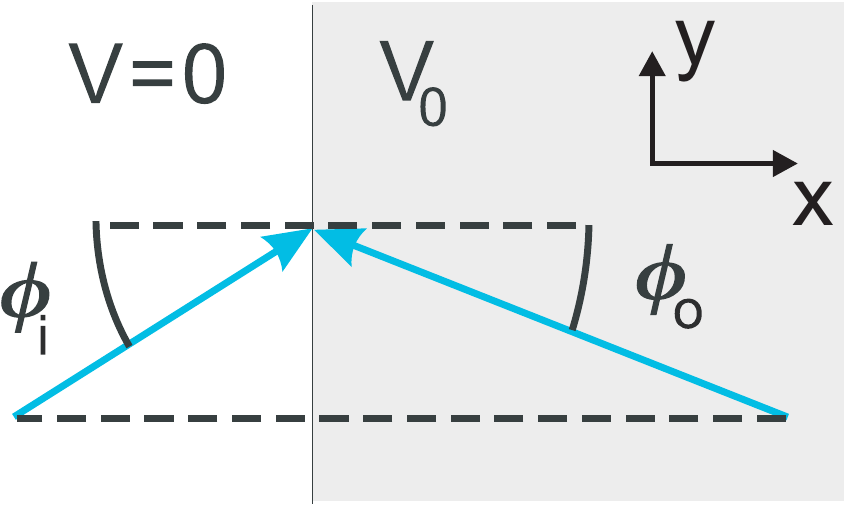}}
    \end{center}
  \caption{Left: A 1D potential barrier of height $V_0$ and width $W$. 
  	Right: wave vector $\vec{k}=(k_x,k_y)$ geometry for an electron, with energy 
  	$0 < E < V_0$, impinging on a square potential barrier (grey area).
    }\label{fig_transdelta1}
\end{figure}
Referring to Eq. (\ref{eq1_2}) and Fig. \ref{fig_transdelta1}, the wave function 
in each of the  regions (1)-(3) can be written as 
a superposition of the eigenstates of Eq. (\ref{eq1_2}) 
$\psi_n(x) = \mathcal{G}_n \mathcal{M}_n(x)\mathcal{A}_n, n=1,2,3$, with 
coefficients 
$\mathcal{A} = (A,B)^T$ and
\begin{equation}\label{eq2_2}
\mathcal{G} = \matt{1}{1}{s e^{i \phi}}{-s e^{-i \phi}}\,, 
	\quad  \mathcal{M}(x) = \matt{e^{i \lambda x}}{0}{0}{e^{-i \lambda x}}\,.
\end{equation}
Matching the wave function at the interfaces $x=0$ and $x=W$ gives the transfer 
matrix $\mathcal{N} = \mathcal{G}^{-1} \mathcal{S}\mathcal{G}$ in terms of the 
matrix $S$ which relates the wave function in front of the barrier to the one 
after it in the manner $\psi_1(0) = S \psi_3(0+)$. The result is
\begin{equation}\label{eq2_7}
    S = \mathcal{G}_2 \mathcal{M}^{-1}_2(W) \mathcal{G}^{-1}_2 = 
    \matt{\cos P}{i \sin P}{i \sin P}{\cos P}\,.
\end{equation}
We notice that $S$ is a periodic function of $P$ and that $S = \pm \mathbb{1}$  
for  $P = n \pi$. This is a special situation in which the two pseudo-spin 
components of the wave function do not mix. 
Later on we will see that this periodicity appears 
in the transmission through a $\delta$-function barrier and in the dispersion 
relation of the KP model.

With  the elements of $\mathcal{N}$ denoted by $n_{ij}$, the transmission is 
$T = |t|^2 =  1/|n_{11}|^2$. The explicit result is
\begin{equation}\label{eq2_12}
    T = 1/[1 + \sin^2 P \tan^2 \phi]\,,
\end{equation}
and coincides with the formula for transmission, found in Ref. \cite{kat}, in the limit of $\delta$-function barriers. 
Obviously, $T$ and $R = 1 - T$ are periodic functions of $P$, that is, 
$X(P+n\pi,\phi) = X(P,\phi)$ for $n$ integer and $X=T,R$. In addition, from 
Eq.~(\ref{eq2_12}) we find that  $T(P,\phi)$ has the following properties:
\begin{eqnarray}
\nonumber
\hspace*{-.59cm}& 1) & T(P, \phi) = T(\pi - P, \phi) = T(\pi + P, \phi),\\*
\nonumber
\hspace*{-.59cm}& 2) & T(n\pi, \phi) = 1\,,\\*
\nonumber
\hspace*{-.59cm}& 3) & T(\pi/2, \phi) = \cos^2 \phi\,,\\*
\nonumber
\hspace*{-.59cm}& 4) & T(P,\phi = 0) = 1,\,T \approx 1 \text{ for } \phi 
\approx 0
\leftrightarrow k_y \ll k_x\,, \\*
\hspace*{-0.4cm}& 5) & T(P,\pm \pi/2) = 
0, \, T \approx 0 \text{ for } \phi 
\approx \pm \pi/2 \leftrightarrow k_y \gg k_x\,.
\end{eqnarray}
These results  are very different from those of the non-relativistic case where 
T is a decreasing function of P. A contour plot of the transmission is shown in 
Fig. 2(a). This figure shows clearly the symmetry properties 
$T(P,\phi) = T(P,-\phi)$, and $T(\pi-P,\phi) = T(P,\phi)$.

\subsection{Conductance}
The two-terminal conductance is 
$G=G_0\int_{-\pi/2}^{\pi/2} 
T(P,\phi) \cos\phi\, \mathrm{d}\phi$, with 
$G_0= 2 E_F L_y e^2/ (v_F h^2)$ and $L_y$ the width of the system. 
Using Eq. (4) for 
$T(P,\phi)$ the resulting $G$ is periodic in $P$ and given by
\begin{equation}\label{eq2_14}
    G/G_0 = 2\big[ 1 - \text{artanh}(\cos P) \sin P \tan P\big]/\cos^2 P\,.
\end{equation}
For one period, $G$ is shown in Fig. 2(b); its minimum value is $4/3$ and its 
maximum one $2$.
\begin{figure}[ht]
    \begin{center}
    \subfigure{\includegraphics[height=3.3cm]{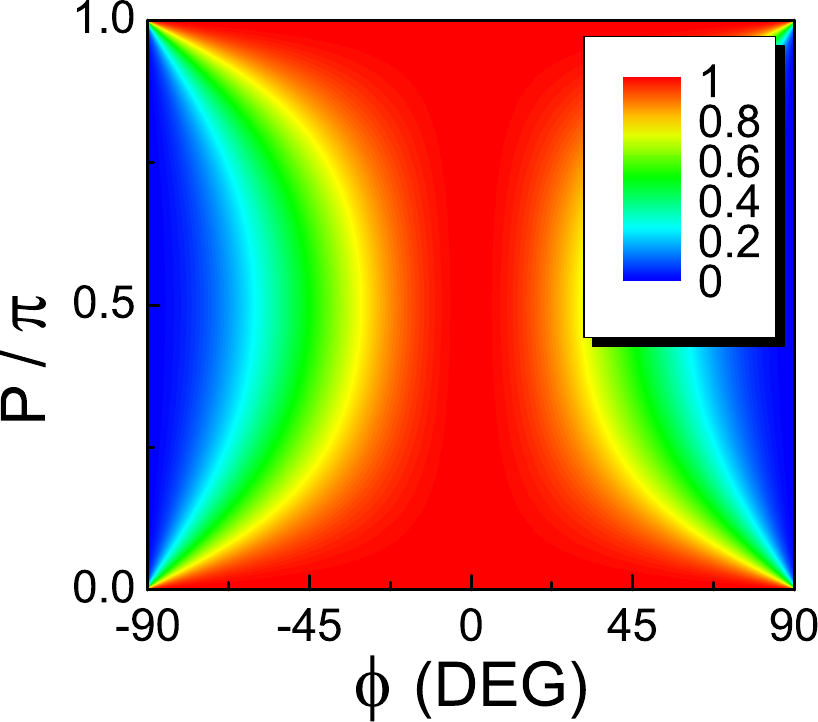}}
    \subfigure{\includegraphics[height=3.3cm]{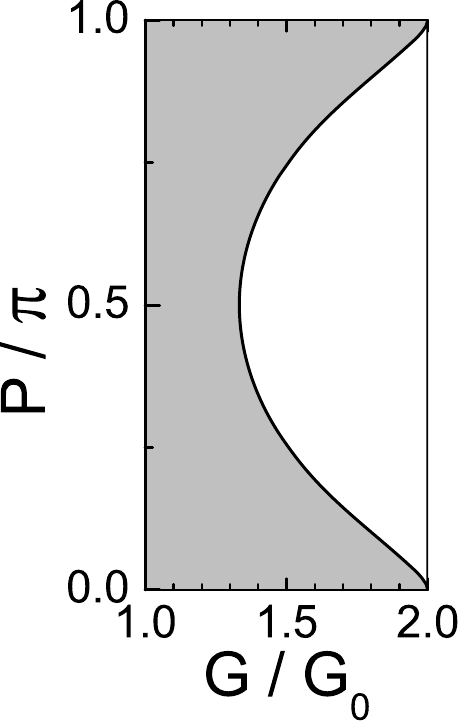}}
    \end{center}
    \caption{ Left: Transmission $T$ through a
    $\delta$-function barrier vs its
    strength
    $P$ and the angle of incidence $\phi$ ($\tan\phi=k_y/k_x$).
    Only one period is shown. Right: Conductance $G/G_0$ vs
    $P$. }
    \label{fig_trans_Pfi}
\end{figure}

\section{Transmission through two $\delta$-function barriers}
We consider two barriers separated by a distance $L$ 
characterized by the potential $V(x,y) / \hbar v_F = P_1\delta(x) + P_2\delta(x - L)$, with strengths $P_{1,2}$ 
and introduce the 
dimensionless variables $\ve \rightarrow \ve L$, $k_y \rightarrow k_y L$, $u_0 
\rightarrow u_0 L$, and $x \rightarrow x / L$. Due to space limitations we 
treat only the cases of parallel and antiparallel $\delta$-function barriers 
with the same strength $ |P_1| = |P_2|$.

{\it Parallel $\delta$-function barriers.}
This is a model system for a resonant tunneling 
structure \cite{kat,per} and 
also for a Fabry-Perot interferometer whose resonances were recently 
investigated experimentally \cite{cho}. The transmission is given by
\begin{equation}\label{eq3_5}
    T = \big[1+ \tan^2\phi ( \cos k_x \sin2 P - 2 s \sin k_x \sin^2 P 
    /\cos\phi)^2 \big]^{-1}\,,
\end{equation}
with $s = sign(\ve)$. The properties of $T(P,\phi,k_x)$ are  identical to those 
for a single barrier except for property 3) and property 1) which must be 
replaced by $T(P, \phi, k_x) = T(P + n \pi, \phi, k_x)$. In Fig. 3(a) we show 
$T(P,\phi,k_x)$ through two barriers with $P = \pi/10$.
\begin{figure}[ht]
    \begin{center}
	\subfigure{\includegraphics[width=
	3.7cm]{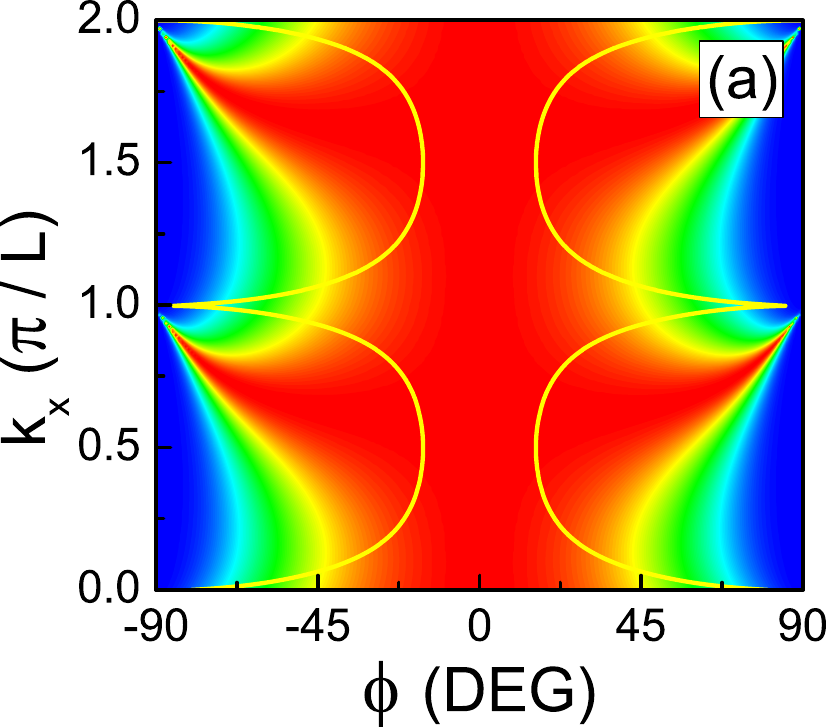}}
	\subfigure{\includegraphics[width=
	3.7cm]{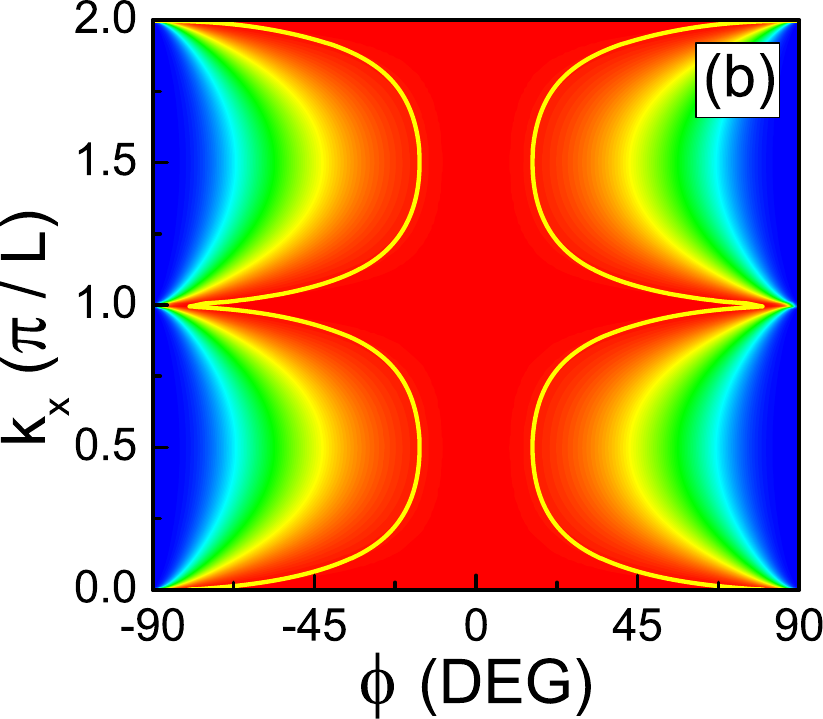}}
	\subfigure{\includegraphics[width=0.7cm ]{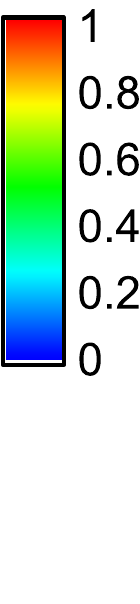}}
    \end{center}
    \caption{(a) Transmission through two {\it parallel} $\delta$-function 
    barriers, as a function of the wave vector component $k_x$ and angle of 
    incidence $\phi$, for $P = \pi/10$. The yellow solid curve shows the contour 
    with a transmission value of 0.75 for $P = \pi/2$. 
    (b) As in (a) for two {\it antiparallel} $\delta$-function barriers.}
\label{fig_trans_kxfi}
\end{figure}

{\it Antiparallel $\delta$-function barriers.} 
We now consider two parallel $\delta$-functions but with opposite sign, 
$P=P_1=-P_2$. The result for the transmission is
\begin{equation}\label{eq3_6}
\hspace*{-0.23cm} T = \big[ \cos^2 k_x + \sin^2 k_x (1 - \sin^2 \phi 
\cos 2 P)^{2}/\cos^4 \phi  \big]^{-1}\,.
\end{equation}
In Fig. 3(b) we show the transmission through two opposite barriers for 
$P = \pi/10$. The symmetry properties of $T(P,\phi,k_x)$ for a single barrier 
again hold, except for the value of $T(\pi/2,\phi,k_x)$, see property 3). In 
addition, we now have $T(P,\phi,k_x) = T(P,\phi,-k_x)$.

The periodicity in the transmission is also present in the conductance $G$. We 
show $G$ in Fig. 4(a) for parallel and in Fig. 4(b) for antiparallel 
$\delta$-function barriers.
\begin{figure}[ht]
    \begin{center}
    \subfigure{\includegraphics[height=3.1cm]{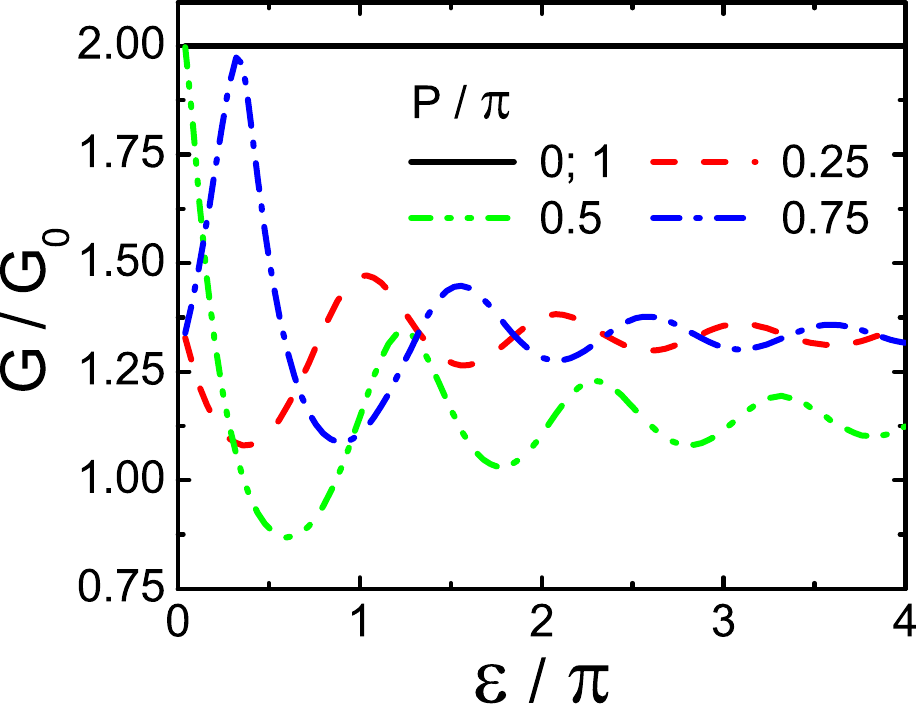}}
    \subfigure{\includegraphics[height=3.1cm]{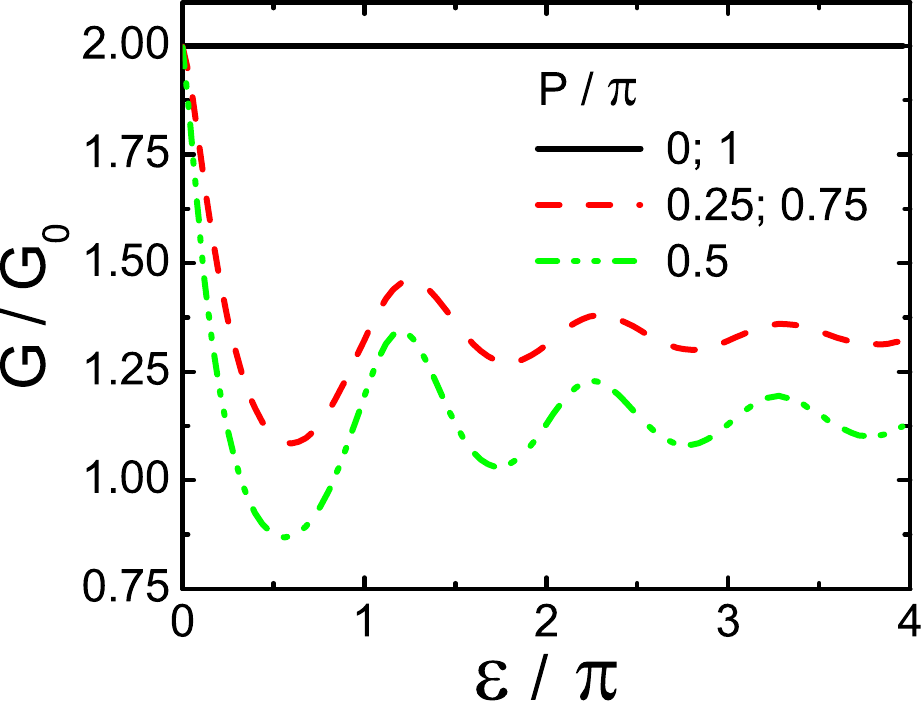}}
    \end{center}
    \caption{(a) Conductance $G(G_0)$ vs 
    $\ve$ for several strengths 
    $P$ through two parallel 
    $\delta$-function barriers. (b) As in (a) for two antiparallel barriers.}
    \label{fig_cond_slice}
\end{figure}

\section{Kronig-Penney model}
For an infinite number of periodic square barriers, one can tune the system into 
a self-collimating material \cite{parksgs}. For special values of $V_0$, $W$, 
and $L$ it was found in Ref.~\cite{parksgs} that the dispersion relation 
near the K point is almost linear in $k_x$ and constant along $k_y$. The system 
thus behaves as a 1D metal. We look for similar results using $\delta$-function 
barriers. Within the KP model we replace the square barriers by 
$\delta$-function barriers, 
characterized by $V(x,y)/\hbar v_F = \sum_{j=-\infty}^{\infty} P \delta(x - j L)$. 
The resulting wave function is a Bloch function and 
the transfer matrix $\mathcal{N}$ pertinent to these  barriers leads to $\psi(1) 
= e^{i k_x} \psi(0)$ and $\mathcal{A}_1 = \mathcal{N} \mathcal{A}_2$, with $k_x$ 
the Bloch wave vector. From these conditions we can extract the relation 
$e^{-i k_x } \mathcal{M}(1) \mathcal{A}_2 = \mathcal{N} \mathcal{A}_2$, with 
$\mathcal{M}(x)$ given by Eq. (\ref{eq2_2}). Then setting the determinant of the 
coefficients in $\mathcal{A}_2 = (A,B)^T$ equal to zero and using the 
transfer matrix for a $\delta$-function barrier leads to 
($\lambda= [\ve^2 - k_y^2]^{1/2}$)
\begin{equation}\label{eq5_5}
    \cos k_x  =  \cos P \cos \lambda + (\ve/\lambda) \sin \lambda \sin P\,.
\end{equation}
The solution of Eq. (\ref{eq5_5}) gives the dispersion which is periodic in $P$ 
and the spectrum 
is shown in Fig. 5 for $P = \pi/2$. Further,  Eq. (\ref{eq5_5}) is mapped, 
for $k_y=0$, directly onto that for strictly 1D fermions\cite{mak} and gives the 
spectrum
\begin{equation}\label{eq5_7}
    \ve = P \pm k_x + 2 n \pi\,,
\end{equation}
with n an integer.

Equation (9) contrasts very 
sharply with that for 2D electrons with a {\it parabolic} 
spectrum in a 1D KP potential which, with $\lambda'= [2\mu \ve - k_y^2]^{1/2}$ 
and 
$\mu = m v_F L / \hbar$, reads
\begin{equation}\label{eq5_6}
    \cos k_x  = \cos \lambda' + (\mu P/\lambda') \sin \lambda'\,;
\end{equation}
the resulting dispersion relation is not periodic in $P$.
\begin{figure}[ht]
\begin{minipage}{0.60\linewidth}
\centering
\includegraphics[width=0.95\linewidth]{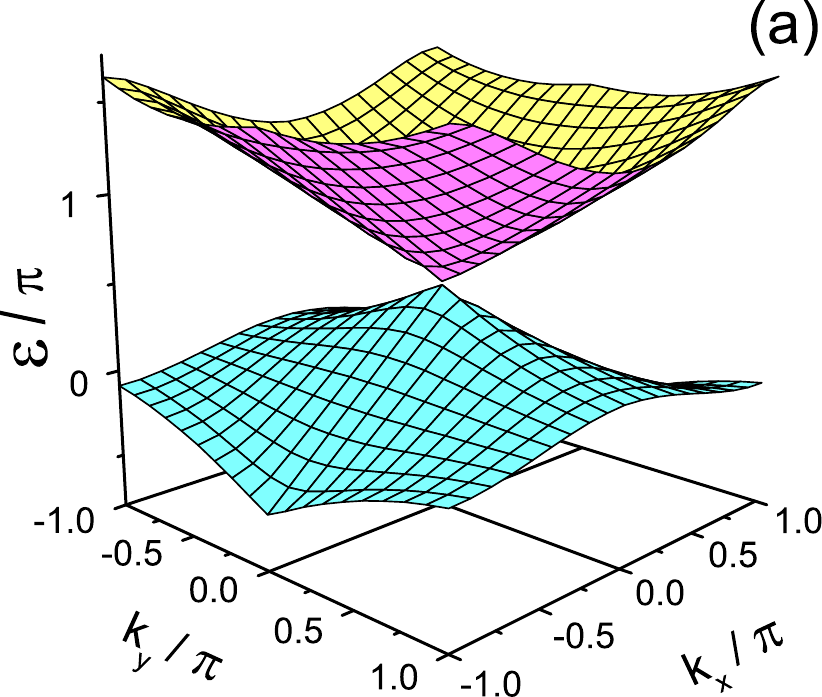}
\end{minipage}
\begin{minipage}{0.35\linewidth}
\centering
\subfigure{\includegraphics[width=1\linewidth]{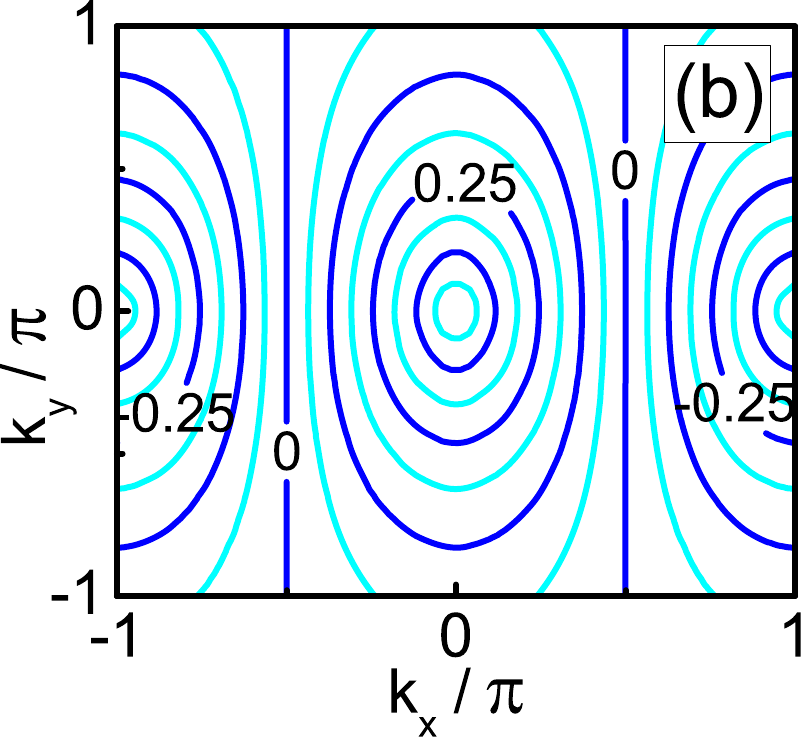}}
\subfigure{\includegraphics[width=1\linewidth]{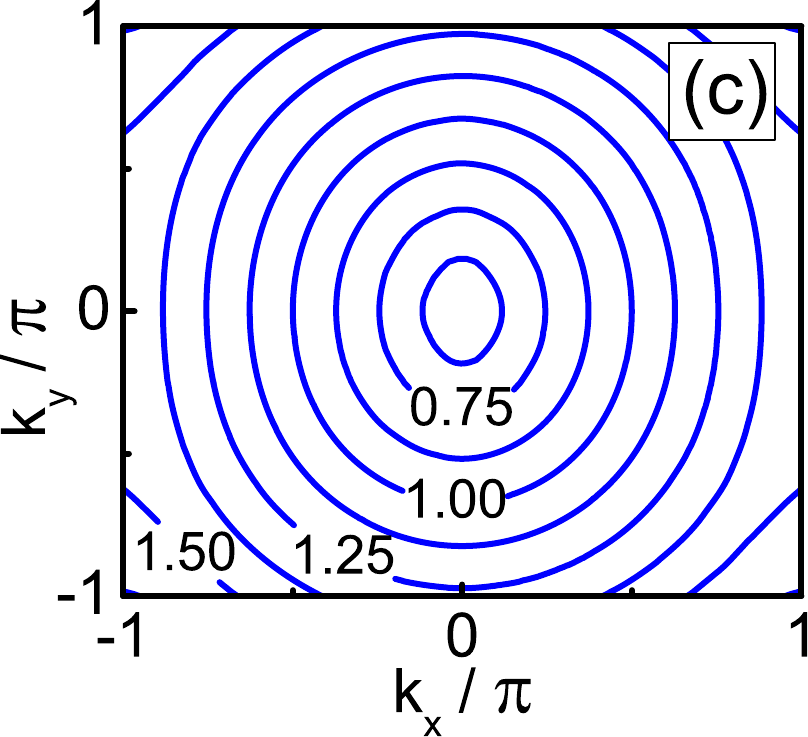}}
\end{minipage}
    \caption{(a): Energy bands, near the Fermi level, close to the K point in 
    the KP model, for $P = \pi/2$. The valence band touches the conduction band 
    at $\ve = P = \pi/2$. For large $k_y$ the valence band becomes flat. (b) and 
    (c): projections, respectively, of the valence  and conduction bands onto 
    the $(k_x,k_y)$ plane.}
    \label{fig_disp3d_KP}
\end{figure}

\subsection{Properties of the spectrum}
Since the dispersion relation is periodic in $P$, with period $2 \pi$, it is 
sufficient to study it only for $0\leq P \leq 2 \pi$.  For particular values of 
$P$ we find:
\begin{eqnarray}
\nonumber
\hspace*{-0.4cm} &&1)\,\,P = 2 \pi\rightarrow\ve = 
\pm [k_y^2 + (k_x + 2 n \pi)^2]^{1/2}\,,\\
\nonumber
&&2)\,\,P = \pi \rightarrow\ve = \pm [k_y^2 + (k_x + (2 n + 1) \pi)^2]^{1/2}\,, \\
&&3)\,\,P = \pi/2 (3\pi/2)\rightarrow\cos k_x = +(-)\ve\sin \lambda/\lambda\,.
\end{eqnarray}

In limiting cases we are able to obtain explicit expressions for 
$E = E(k_x,k_y)$. We expand the dispersion relation for small $k_y$ and 
$\ve- P$. The resulting  quadratic equation for $\ve$ is solved by
\begin{equation}\label{eq5_9}
    \ve_\pm \approx P \pm \big[4 \sin^2 (k_x/2) +(k_y^2/P^2) 
    \sin^2 P\big]^{1/2}\,.
\end{equation}
For small $k_x$ we can replace the term $4 \sin^2 (k_x/2)$ by $k_x^2$. Notice 
that for $k_x = 0$ we find $\ve_\pm \approx P \pm k_y \sin P / P$ which is a 
linear spectrum with a reduced velocity. For $k_y = 0$, we have 
$\ve_\pm = P \pm 2 \sin (|k_x|/2)$, which is linear for small $k_x$ but posseses 
a typical band shape for large $k_x \approx \pi$. For small 
$k_x > k_y \sin P /P$ we have
\begin{equation}\label{eq5_10}
\ve_\pm \approx P \pm \Big(2k_x^2 + k_y^2 \sin^2 P/P^2\Big)\Big/2 |k_x|\,.
\end{equation}
For $P\gg 1$, $\ve$ is highly anisotropic and nearly flat vs $k_y$.

{\it Relation to the spectrum of a square superlattice.}
We now  look whether we can find an energy spectrum similar to that of Ref. 
\cite{parksgs} pertinent to square barriers, with height  $V_0 = 720$ meV, 
and width $w = 5$ nm,  and  unit-cell  length $L = 10$ nm. In our units these 
values correspond to $P = V_0 W / v_F \hbar = 2 \pi$ and lead to: 
$\ve = \pm [k_y^2 + (k_x + 2 n \pi)^2]^{1/2}$. Since the Fermi level in these 
units is $\ve_F = 2 \pi =P$, we look for the spectrum near the  value 
$\ve = [k_y^2 + (\pm |k_x| + 2 \pi)^2]^{1/2}$. Although these bands seem to 
fullfil our demands because the dispersion looks rather flat in the $k_y$ 
direction, the concern is that we would obtain the same dispersion for  $P\to 0$ 
and $\ve_F\to 2 \pi$. But this can be obtained by folding the cone-like 
dispersion of graphene and results simply from working in the reduced-zone 
scheme. Consequently  no new fundamental physics should be attached to it. 
Further, from this correspondence we expect and found that for square barriers 
with  $P=2 \pi n$, the situation is more favorable  for the occurrence of 
collimation. It follows that the collimation effect is also obtainable for 
barriers that are lower than the unusually high ones of Ref. 
\cite{parksgs} if one uses longer unit-cell periods.

\section{Extended Kronig-Penney model}
The square barriers are replaced by alternating-in-sign $\delta$-function 
barriers. The unit cell of the periodic potential contains one such barrier up, 
at $x = 0$, followed by a barrier down, at $x = 1/2$. The resulting transfer 
matrix leads to
\begin{equation}\label{eq5_11}
    \cos k_x = \cos \lambda - (2 k_y^2/\lambda^2) \sin^2 (\lambda/2) \sin^2 P\,,
\end{equation}
where  $\tan \phi = k_y / \lambda$. From Eq. (15) we deduce that the dispersion 
is periodic in $P$, with period $\pi$, and has the following properties:
\begin{eqnarray}
\nonumber
    \hspace*{-0.5cm}&&1) \text{ it is invariant for } \ve \rightarrow -\ve 
    \text{ and } P \rightarrow \pi - P\,,\\
\nonumber
    \hspace*{-0.5cm}    && 2)\,\, P = n \pi \rightarrow\ve = 
    \pm [k_y^2 + (k_x + 2 n \pi)^2]^{1/2},\\
    \hspace*{-0.5cm}    &&3)\,\,P = \pi/2\rightarrow (\ve,k_x,k_y) = (0,0,k_y).
\end{eqnarray}
In Fig. 6(a) we show the spectrum for $P = \pi/2$. As seen, it is almost 
independent of $k_y$ for small energies  while in the $k_x$ direction the bands 
are linear; this is an advantageous situation for  self-collimation.
\begin{figure}[ht]
    \begin{center}
    \subfigure{\includegraphics[height=4cm, 
    width=4cm]{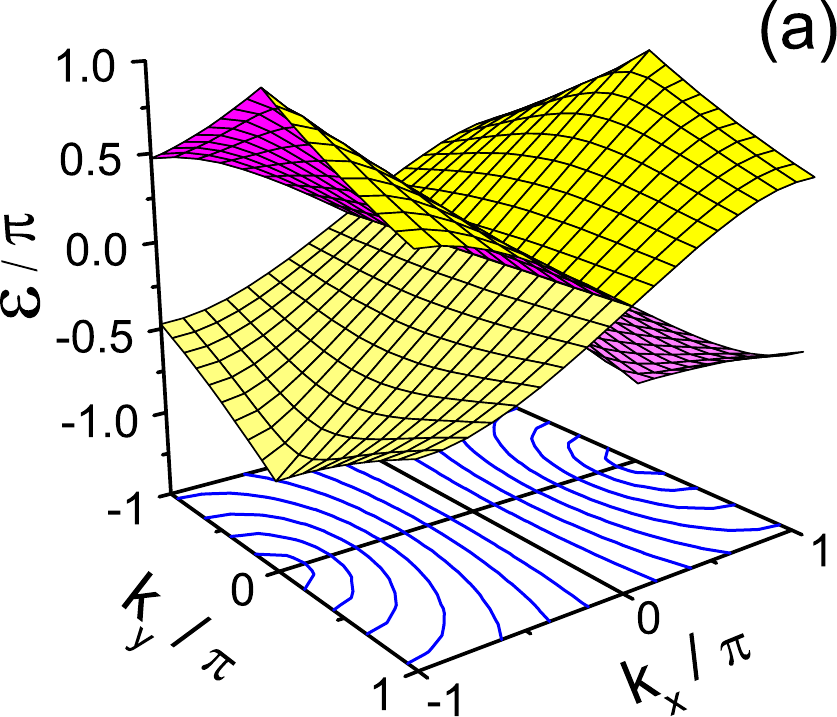}}
    \subfigure{\includegraphics[height=4cm, 
    width=4cm]{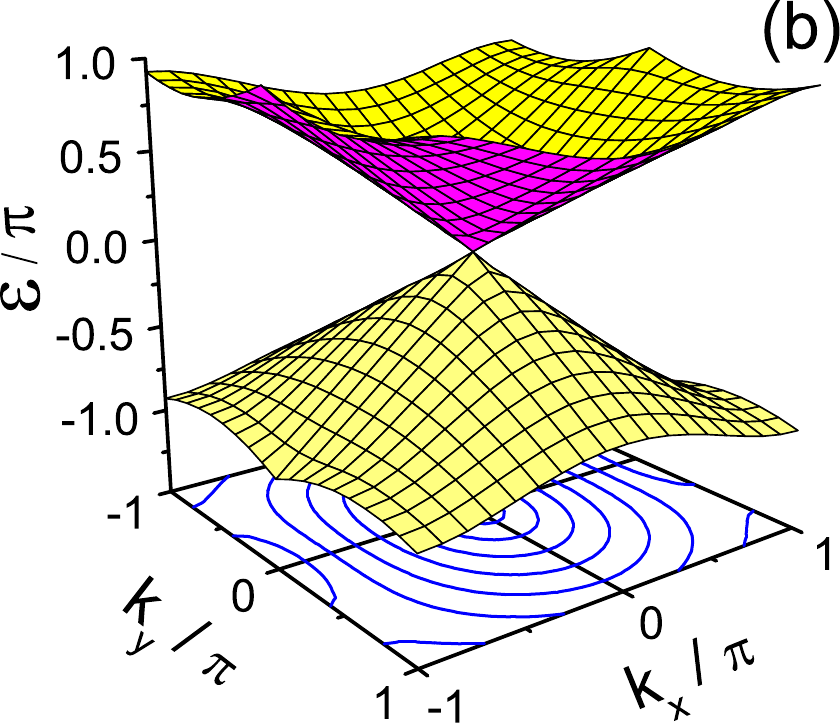}}
    \end{center}
    \caption{(Color online) (a) Conduction and valence bands, near the K point, 
    in the extended KP model, for $P = \pi/2$. The bands form a ``$ $cross'' in 
    the $(E, k_x)$ plane and the Dirac point has become a Dirac line. The 
    dispersion is  nearly independent of  $k_y$. (b) As in (a) for 
    $P = \pi / 4$.}
    \label{fig_disp3d_KPext}
\end{figure}
For $k_y = 0$ we obtain the linear spectrum
\begin{equation}\label{eq5_12}
    \ve = \pm |k_x| + 2 n \pi\,,
\end{equation}
with the Dirac point at $\ve = 0$. We can also find an explicit expression for 
$k_x \approx 0$. Solving Eq. (\ref{eq5_11}) gives
\begin{equation}\label{eq5_14}
    \ve_\pm = \pm |k_y \cos P|\,.
\end{equation}
Then the group velocity $v_y \propto \partial \ve / \partial k_y$ becomes small 
if $P \approx \pi/2 + n \pi$. 
Fig. 6(b) shows the energy spectrum for $P = \pi/4$, the Dirac cone becomes 
anisotropic, as the spectrum flattens in the $k_y$ direction.

We now consider the case where $k_x$ and $k_y$ are nonzero. 
If $\ve \ll 1$ then the rhs of (\ref{eq5_11}) can be expanded in $\ve$. 
This leads to a quadratic equation for $\ve$ with solutions
\begin{eqnarray}\label{eq5_17}
\nonumber
     \ve& \approx& \pm |k_y|\left[\frac{\cosh k_y-\cos k_x  - f
     }{(k_y/2)\cos^2 P \sinh k_y + f
     }\right]^{1/2} \\*
    && \xrightarrow{P=\pi/2} \pm |k_y|
    \sin (|k_x|/2)/\sinh (|k_y|/2)\,,
\end{eqnarray}
where $f= 2 \sin^2 P \sinh^2 (k_y/2)$. For $k_y = 0$ we find the  result given 
by Eq. (\ref{eq5_9}), that is, $\ve_\pm = \pm 2 \sin(|k_x|/2)$ which is linear 
for small $k_x$.

\section{Conclusions}
In summary, we studied the  transmission and conductance of fermions, with 
energy {\it linear in wave vector}, through one and two $\delta$-function 
barriers  and the energy spectrum of a KP superlattice. 
For very high ($V_0\to \infty$) and very thin ($W \to 0$)  barriers we showed 
that they are {\it periodic} functions of their strength $P = W V_0/ \hbar v_F$ 
where $v_F$ is the Fermi velocity. 
Further, we showed that a KP superlattice has an energy spectrum that is a 
{\it periodic} function of $P$, which is in sharp contrast with that obtained 
from the Schr\"odinger equation. An important consequence of that is collimation 
of an incident electron beam\cite{parksgs} that here occurs for $P=2\pi n$ with 
$n$ an integer. We also obtained various explicit but approximate dispersion 
relations, e.g., for small wave vector $\vec{k} = (k_x,k_y)$. Given the intense 
research activity in graphene and the very recent experimental verification of 
Klein tunneling\cite{kex}, we expect that this {\it periodic} dependence on 
the strength $P$ will be tested in a near future.

\acknowledgments
This work was supported by IMEC,
the Flemish Science Foundation (FWO-Vl), the Belgian Science Policy
(IAP), and the Canadian
NSERC Grant No. OGP0121756.
\end{document}